\documentclass[final,3p]{elsarticle}

\usepackage{times}

\usepackage{setspace}
\usepackage{amssymb}
\usepackage{amsmath}

\newcommand{\incfig}{\centering\includegraphics}
\setkeys{Gin}{width=0.7\linewidth,keepaspectratio}

\newcommand{\eqr}[1]{Eq.\thinspace(#1)}

\newcommand{\ignore}[1]{}  

\journal{Journal of Computational Physics}

\begin{document}

\begin{frontmatter}



\title{On discontinuous Galerkin discretizations of second-order derivatives}%


\author[pppl,mppc]{A.~H. Hakim}%
\author[pppl]{G.~W. Hammett}%
\author[pu]{E.~L. Shi}%

\address[pppl]{Princeton Plasma Physics Laboratory, Princeton, NJ
  08543-0451}%
\address[mppc]{Max-Planck/Princeton Center for Plasma Physics,
  Princeton University, Princeton, NJ}%
\address[pu]{Department of Astrophysical Sciences,
  Princeton University, Princeton, NJ}%


\begin{abstract}
Some properties of a Local discontinuous Galerkin (LDG) algorithm are
demonstrated for the problem of evaluting a second derivative $g =
f_{xx}$ for a given $f$.  (This is a somewhat unusual problem, but it is
useful for understanding the initial transient response of an algorithm
for diffusion equations.)  LDG uses an auxiliary variable to break this
up into two first order equations and then applies techniques by analogy
to DG algorithms for advection algorithms.  This introduces an asymmetry
into the solution that depends on the choice of upwind directions for
these two first order equations.  When using piecewise linear basis
functions, this LDG solution $g_h$ is shown not to converge in an
$L_2$ norm because the slopes in each cell diverge.  However, when LDG
is used in a time-dependent diffusion problem, this error in the second
derivative term is transient and rapidly decays away, so that the
overall error is bounded.  I.e., the LDG approximation $f_h(x,t)$ for a
diffusion equation $\partial f / \partial t = f_{xx}$ converges to the
proper solution (as has been shown before), even though the initial rate
of change $\partial f_h / \partial t$ does not converge.  We also show
results from the Recovery discontinuous Galerkin (RDG) approach, which
gives symmetric solutions that can have higher rates of convergence for
a stencil that couples the same number of cells.
\end{abstract}

\end{frontmatter}

\section{Introduction}

Discontinuous Galerkin (DG) algorithms have been widely studied and
fruitfully applied to a wide range of problems in recent
years\cite{Cockburn:1998vt,Cockburn:2005,Dawson:2006}.  Here we focus on
the problem of discretizing the second derivative of a known function.
There are certain subtleties about the behavior of some algorithms for
this problem, and understanding these can be helpful for understanding
aspects of the algorithms for other problems, such as diffusion or
elliptic problems.

Consider the problem of computing the discrete second derivative of a
function, $f(x)$, given its projection $f_h(x)$ on a piecewise
discontinuous polynomial space. Denoting $g(x) = f_{xx}$, the problem
is to determine the best (in some $L_2$ sense) representation $g_h(x)$
on the same space. Discretization of such operators is required in a
large number of problems involving diffusive processes. Our particular
interest is to discretize certain forms of collision operators. For
example, for small angle scattering in a plasma, the Lenard-Bernstein
collision operator, $\mathcal{C}_{LB}[F(v)] = \left( \nu (v-u)F + \nu
  v_t^2 F_{v} \right)_{v}$, an approximation to the full Landau
operator, is a combination of velocity space drag and diffusion,
relaxing the distribution function to a Maxwellian. Even in the
absence of an explicit diffusive process, a ``hyper-collision''
operator (involving terms such as $f_{xxxx}$ or higher even-order
derivatives) may be required to prevent momentum space filamentation
and recurrence
problems\cite{Hammett:1992tf,Hammett:1993,Cheng:2013bs}. Subgrid
models also require such operators to account for energy
transfer to unresolved scales.

At present there are three broad techniques for including diffusive
terms. The most popular approach is to introduce auxiliary variables
to rewrite the second-order PDE as a system of first-order PDEs, and
use the DG framework on the resulting larger system. This \emph{local}
DG\cite{Cockburn:1998uy} (LDG) approach leads naturally to estimates
of solution derivatives at cell interfaces by particular choices of
numerical fluxes for the introduced auxiliary variables. The second
technique is to introduce special numerical fluxes, combined with
``penalty'' terms that penalize the solution for being discontinuous
across interfaces\cite{Liu:2010eq}. These two techniques are not
completely independent, and, in some ways, result from an attempt to
apply ideas from finite-element methods to the DG discretization. See
Ref~\cite{Arnold:2002wu} for a review of such methods.

The third technique is to
reconstruct\cite{Huynh:2009hi,vanLeer:2005kk} a continuous
representation of the solution in the two cells sharing an interface
and use that instead to compute the needed derivatives. This
\emph{recovery} DG (RDG) approach is closer in spirit to finite-volume
methods in which solution gradients at cell interfaces, needed in
Navier-Stokes equations, for example, are reconstructed using
least-square fitting.

\begin{figure}[htb]
  \setkeys{Gin}{width=0.45\linewidth,keepaspectratio}
  \incfig{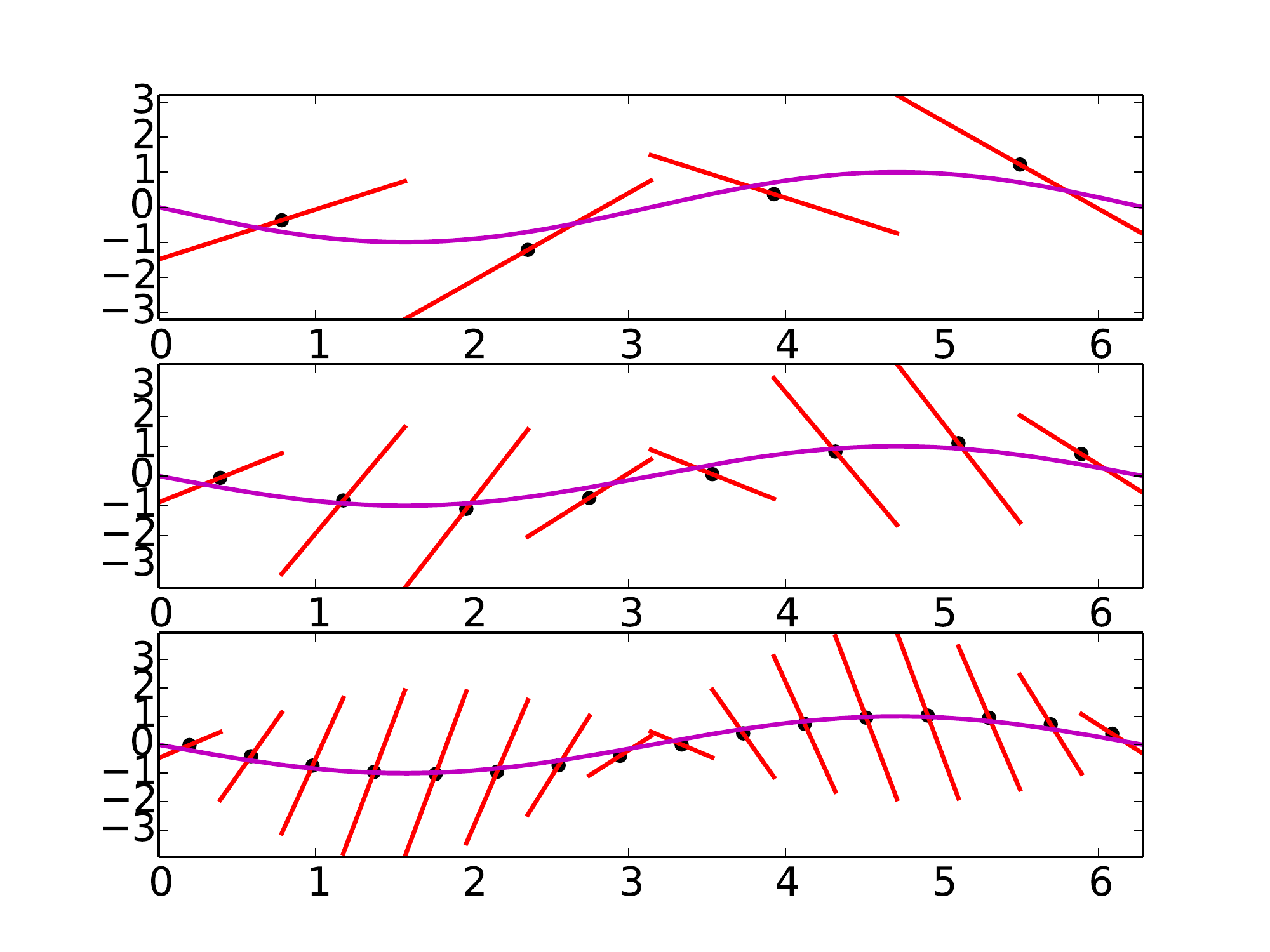}%
  \incfig{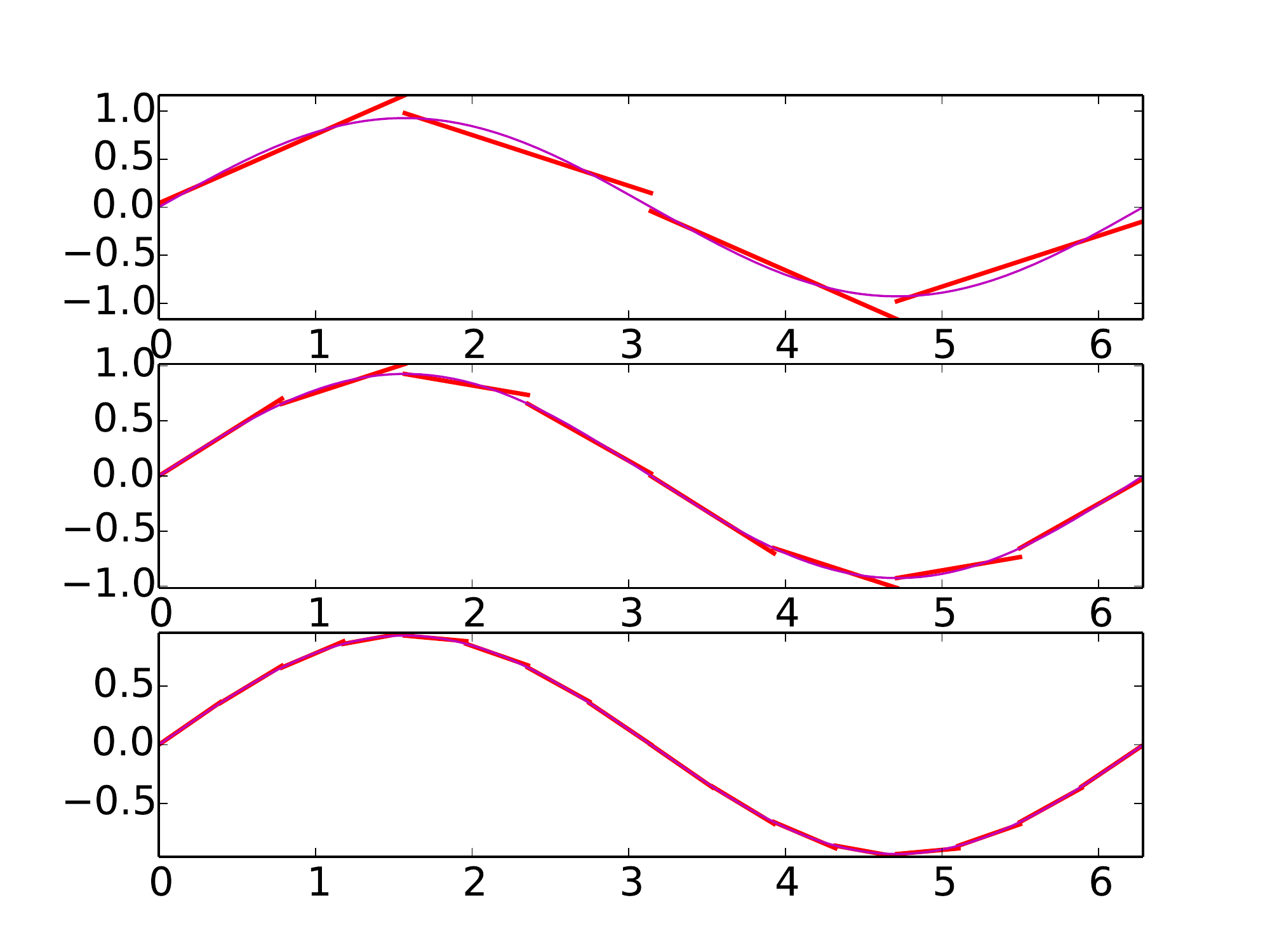}
  \caption{Second derivative of $\sin(x)$ (left panel) with 4 (top), 8
    (middle) cells and 16 (bottom) cells using the asymmetric LDG
    scheme. Solution of diffusion equation $f_t=f_{xx}$ with $\sin(x)$
    initial condition (right panel) with 4 (top), 8 (middle) cells and
    16 (bottom) cells using the same asymmetric LDG stencil. The
    direct computation of second derivatives leads to a converging
    cell average (black dots), but a non-convergent and incorrect
    slope. In the time-dependent case, however, these errors decay
    very rapidly, leading to a convergent scheme. In each test
    $\sin(x)$ is \emph{projected exactly} on linear basis functions.
    The right panels are plotted at $t=0.075$, about the time of
    maximum error for the asymmetric LDG algorithm, as shown in
    Fig.\thinspace{\ref{fig:err-time-hist}}.}%
  \label{fig:ldg-grid-conv}
\end{figure}

\begin{figure}[htb]
  \setkeys{Gin}{width=0.45\linewidth,keepaspectratio}
  \incfig{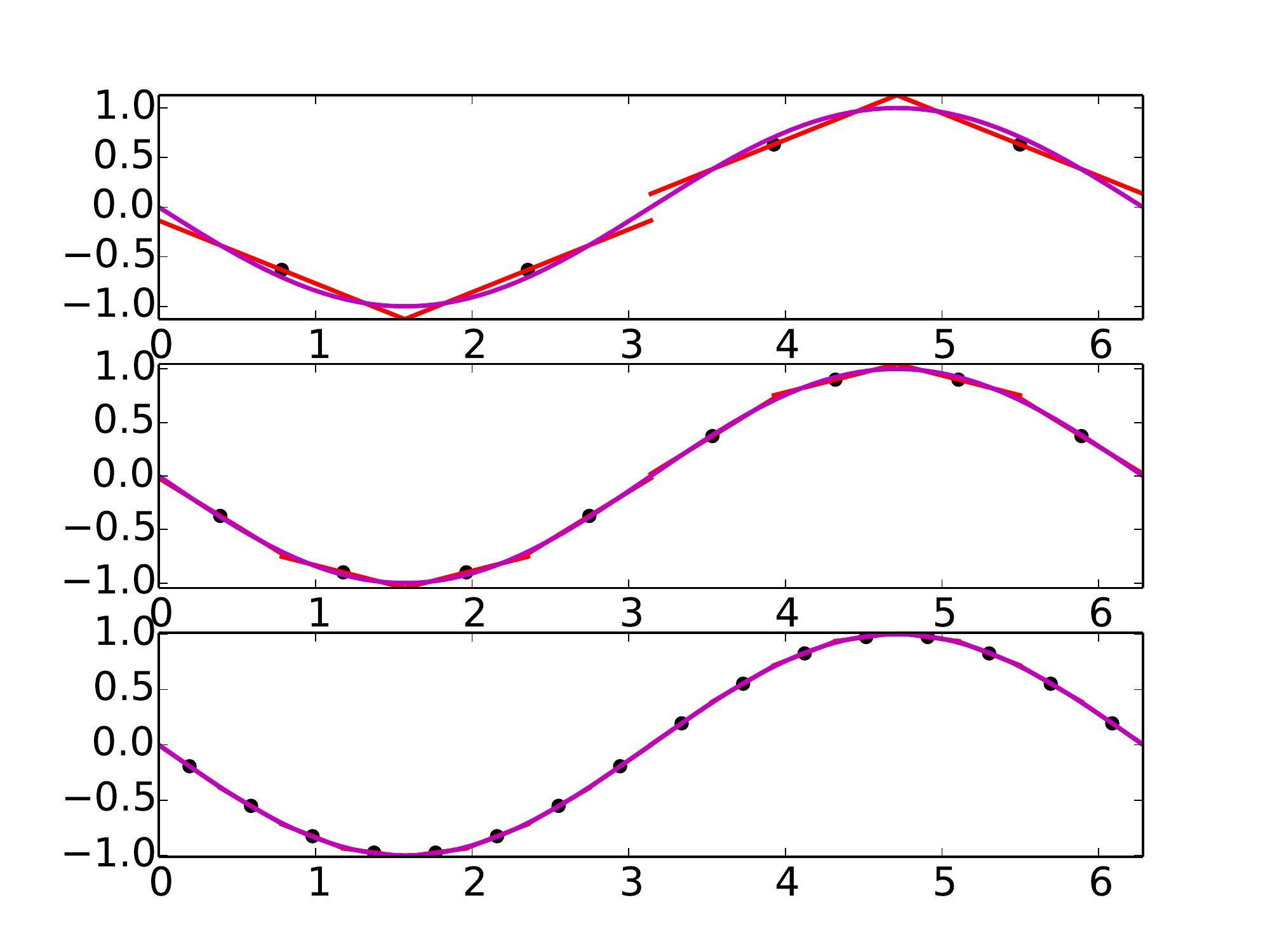}%
  \incfig{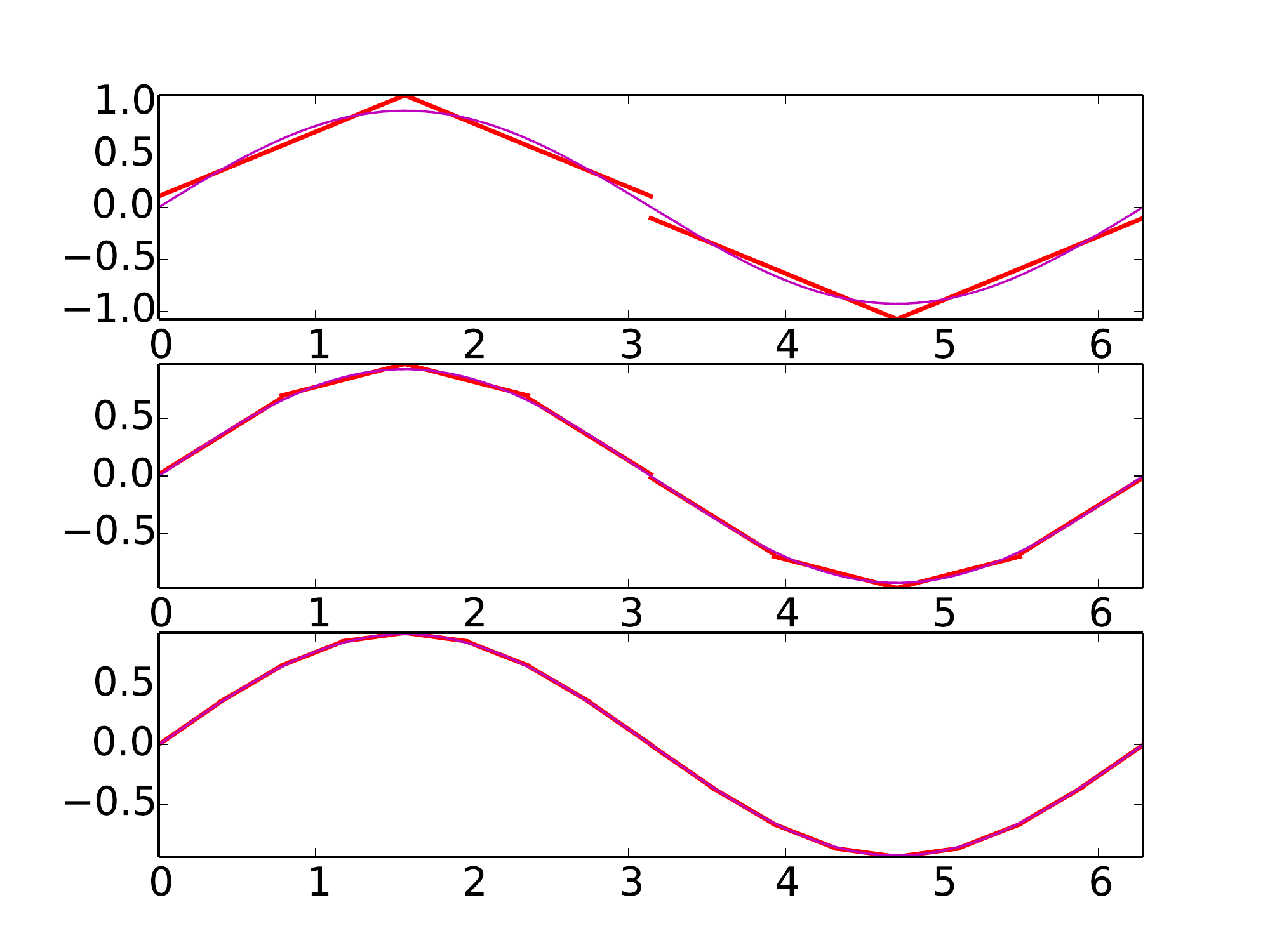}
  \caption{Same as Fig.\thinspace(\ref{fig:ldg-grid-conv}), except
    using the RDG scheme. The cell average as well as the cell slope
    converge rapidly for computing the second derivatives as well as
    in the time-dependent problem. As seen from
    Fig.\thinspace(\ref{fig:err-time-hist}), the errors in the RDG are
    smaller than in either of the LDG schemes. The right panels are
    plotted at $t=0.075$ as in
    Fig.\thinspace{\ref{fig:ldg-grid-conv}}.}%
  \label{fig:rdg-grid-conv}
\end{figure}

\section{Testing discretizations of the second derivative operator}

In this paper we look closely at the LDG and the RDG approaches for
computing second derivatives both directly as well as for solving the
diffusion and Poisson equations. To unify the presentation, assume
piecewise linear expansion $f_h(x,t) = f_{0,j} + f_{1,j}
2(x-x_j)/(\Delta x)$ and $g_h(x,t) = g_{0,j} + g_{1,j}
2(x-x_j)/(\Delta x)$ for $x \in [x_{j-1/2},x_{j+1/2}]$, and where $x_j
= (x_{j-1/2}+x_{j+1/2})/2$. For the LDG scheme rewrite $g=f_{xx}$ as a
system of first order equations, $q = f_x, \; g = q_x$. Multiply by a
test function and integrating over a cell $j$ leads to a weak-form,
which requires determining the values of $f_{j+1/2}$ and $q_{j+1/2}$ at
cell boundaries. In the LDG approach, among several possible choices,
one can pick, for example, $f_{j+1/2}=f_{j+1/2}^-$ and
$q_{j+1/2}=q_{j+1/2}^+$, or $f_{j+1/2}=f_{j+1/2}^+$ and
$q_{j+1/2}=q_{j+1/2}^-$ (these lead to the two forms of the asymmetric
LDG we discuss below).  With Legendre polynomials used as test and
basis functions, for piecewise linear case, evaluating all terms
explicitly, we can show that one form of the LDG scheme can be
expressed as the stencil update
\begin{align}
  \left(
    \begin{array}{cc}
      g_{0,j}\\
      g_{1,j}
    \end{array}
  \right)
  =
  \frac{1}{\Delta x^2}\left(
    \begin{array}{cc}
      4T^{-1} -8I+4 T & 2T^{-1}+2I -4T \\
      -12 T^{-1} +6I + 6T & -6 T^{-1} -24I -6T
    \end{array}\right)\left(\begin{array}{c}
      f_{0,j} \\
      f_{1,j} 
    \end{array}
  \right),
  \label{eq:1}
\end{align}
where the shift operators and its inverse are defined as $T f_{k,j} =
f_{k,j+1}$ and $T^{-1} f_{k,j} = f_{k,j-1}$, and $If_{k,j}=f_{k,j}$,
where $f_{k,j}$ is the $k$-th moment of a modal expansion in Legendre
polynomials in cell $j$.  To derive these (and other) explicit forms
of the update stencils, the auxiliary variable $q$ is eliminated, and,
on use of Legendre polynomial expansion, the stencil obtained. This
stencil corresponds to the LDG method used for the diffusion operator
part of Eq.\ 2.8 of Ref.~\cite{Cockburn:1998uy}, with $\mathbb C$ in
their Eq.\ 2.9 given by $c_{11}=c_{22} = 0$ and $c_{21} = -c_{12} =
\sqrt{a}/2$ (and $a=1$).

Note that the stencil is not symmetric in $j$, i.e., it is not
symmetric with respect to the transformation ($x \rightarrow -x$,
$f_{1,j} \rightarrow -f_{1,j}$, $g_{1,j} \rightarrow -g_{1,j}$), which
corresponds to ($T \leftrightarrow T^{-1}$, $f_{1,j} \rightarrow
-f_{1,j}$, $g_{1,j} \rightarrow -g_{1,j}$) in \eqr{\ref{eq:1}}.
This is a feature of the LDG scheme as one must make a choice of
``upwind'' directions to compute the numerical fluxes for the
first-order system. In fact, one can derive another stencil by
switching the order of the upwind directions, which corresponds to
$c_{21}=-c_{12} 
= -\sqrt{a}/2$. Averaging the two asymmetric stencils, however, leads
to a symmetric LDG stencil
\begin{align}
  \left(
    \begin{array}{cc}
      g_{0,j}\\
      g_{1,j}
    \end{array}
  \right)
  =
  \frac{1}{\Delta x^2}\left(
    \begin{array}{cc}
      4T^{-1} -8I+4 T & 3T^{-1}-3 T \\
      -9 T^{-1} +9 T & -6 T^{-1} -24I-6 T
    \end{array}
  \right)\left(\begin{array}{c}
      f_{0,j} \\
      f_{1,j} 
    \end{array}
  \right).
  \label{eq:s-ldg}
\end{align}
Note that the asymmetric and symmetric stencils have the same terms on
the diagonal, but different off-diagonal terms.

The recovery DG scheme starts from a weak-form. Multiply $g=f_{xx}$ by
a test function $\varphi(x)$ and integrate over a cell to get
\begin{align}
  \int_{I_j} \varphi g dx
  =
  (\varphi f_x - \varphi_x f)\bigg|^{x_{j+1/2}}_{x_{j-1/2}}
  +
  \int_{I_j} \varphi_{xx} f dx,
\end{align}
where integration by parts has been used twice. Note that as the
function $f(x)$ is discontinuous across cell edges, a prescription is
required to compute its value \emph{and} its first derivative at cell
edges.

\begin{figure}[htb]
  \setkeys{Gin}{width=0.75\linewidth,keepaspectratio}
  \incfig{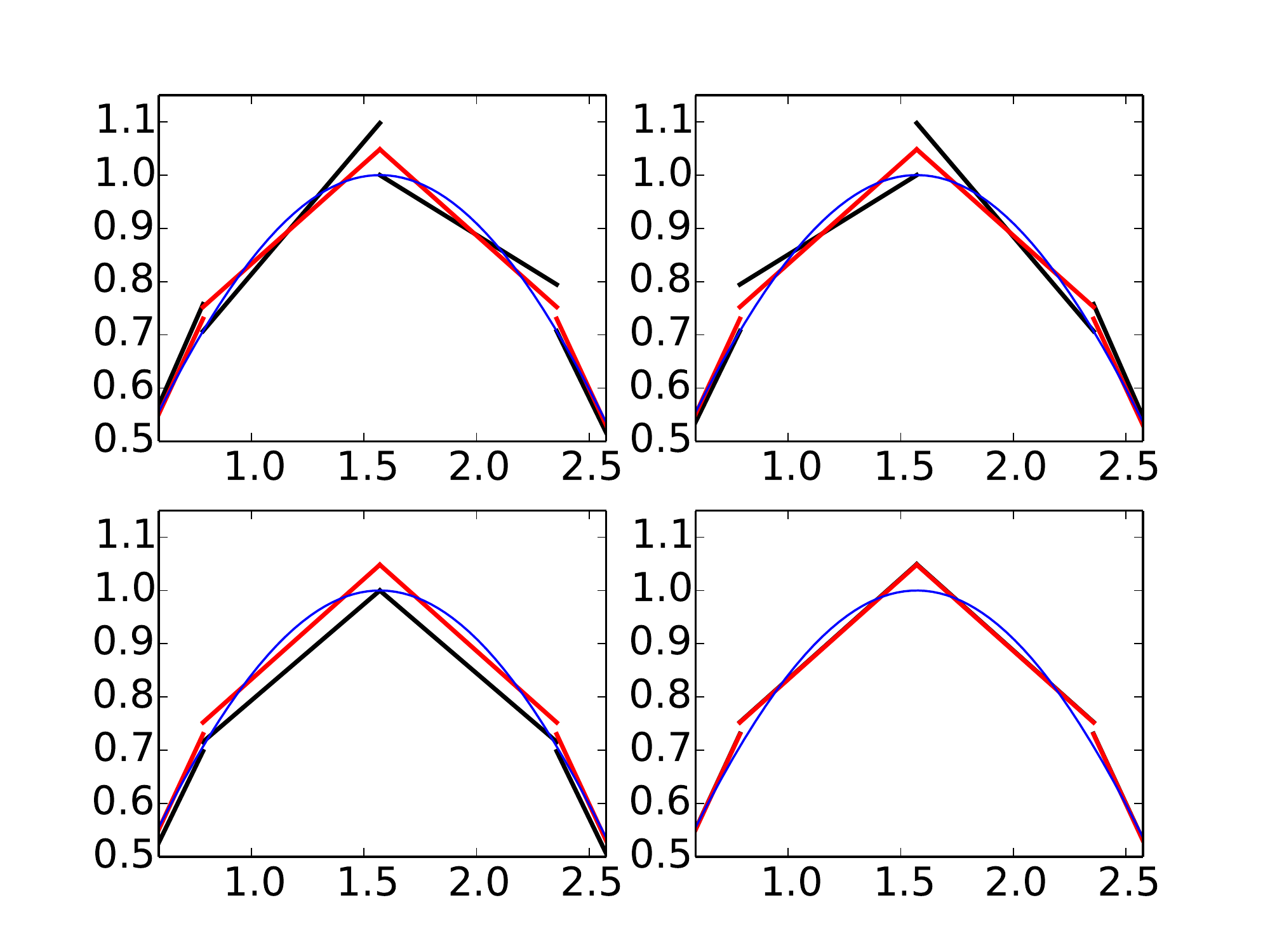}
  \caption{Solution to the Poisson equation $f_{xx}=-\sin(x)$ with
    piecewise linear basis functions on a grid of 8 cells, using the
    two asymmetric versions of LDG schemes (top row), symmetric LDG
    scheme (lower left) and RDG scheme (lower right). Only the
    solution around the maxima are shown.  Black lines are numerical
    solutions and red lines are the projection of the exact solution
    $\sin(x)$ (shown as blue lines) onto piecewise linear basis
    functions. The LDG schemes show larger errors in the slopes than
    the RDG scheme (which matches the exact solutions very
    accurately). However, the LDG computed slopes converge to the
    correct result as the grid is refined.}\label{fig:poisson-sol}
\end{figure}

The recovery discontinuous Galerkin (RDG) scheme proposed by van Leer
and Nomura\cite{vanLeer:2005kk} replaces the function $f(x)$ at
each edge by a \emph{recovered} polynomial that is continuous across
the cells shared by that edge. I.e., we write instead the weak form
\begin{align}
  \int_{I_J} \varphi g dx
  =
  (\varphi \hat{f}_x - \varphi_x \hat{f})\bigg|^{x_{j+1/2}}_{x_{j-1/2}}
  +
  \int_{i_J} \varphi_{xx} f dx,
\end{align}
where $\hat{f}(x)$ is the recovered polynomial, continuous across a
cell edge. As the recovered polynomial is continuous, its derivative
can be computed and used in the above weak-form. Consider an edge
$x_{j-1/2}$. To recover a polynomial that is continuous in cells
$I_{j-1}$ and $I_{j}$ that share this edge, we use $L_2$ minimization
to give the conditions
\begin{align}
 \int_{I_{j-1}} (\hat{f}-f) \varphi_{j-1} dx = 0, \qquad
 \int_{I_{j}} (\hat{f}-f) \varphi_{j} dx = 0
\end{align}
for all test functions in the two cells, $\varphi_{j-1}$ and $\varphi_{j}$.
This ensures that $\hat{f}$, defined over $I_{j-1}\cup I_j$ is
identical to $f(x)$ in the least-square sense, in the space spanned by
the test functions $\varphi_j$.  [For example, in the simplest 
DG case of piecewise constant basis functions, this procedure leads to
an $\hat{f}(x)$ that is a linear function that matches the mean value in
cells $j$ and $j+1$.  For DG with $p$ order basis functions, the full
recovered $\hat{f}(x)$ would be a $2 p +1$ order polynomial, which we
will use here.  One could consider lower order recovery methods
also, where $\hat{f}(x)$ was determined by projection onto a lower order
subset of basis functions.]
For the piecewise
linear case, this procedure leads to the following stencil
\begin{align}
  \left(
    \begin{array}{cc}
      g_{0,j}\\
      g_{1,j}
    \end{array}
  \right)
  =
  \frac{1}{4\Delta x^2}
  \left(
    \begin{array}{cc}
      9T^{-1}-18I+9T & 5T^{-1} + 5T \\
      -15 T^{-1} -15 T & -7 T^{-1} - 46I - 7T
    \end{array}\right)\left(\begin{array}{c}
      f_{0,j} \\
      f_{1,j} 
    \end{array}
  \right).
  \label{eq:rdg}
\end{align}
Note that the stencil is symmetric, which results from the fact that
the recovery procedure does not distinguish the solutions in cells
$j-1$ and $j+1$, as does the LDG scheme.


One way to think about the motivation for the RDG procedure is to
consider that the general DG algorithm tells us how certain moments of
the solution in each cell evolve in time, but there is flexibility in
how to use that information to reconstruct a representation for
$f_h(x)$.  The standard DG approach uses only moment information in a
single cell to construct a representation for $f_h(x)$ within that cell,
which may be discontinuous with the representation in the neighboring
cell.  The RDG algorithm uses information from neighboring cells to
reconstruct a locally continuous representation for $\hat{f}(x)$ to
calculate the flux from the diffusive term, which avoids the problem
that the flux would be infinite if the solution was discontinuous.  In
this regard, RDG is similar in philosophy to the reconstruction approach
described in Ref.~\cite{Dumbser:2008}. For the flux from the advection
operator, one can still use the standard DG approach based on $f_h(x)$
from the upwind side of a cell face, thus preserving the property that
advection should propagate information only in the downwind direction.

\begin{figure}[htb]
  \setkeys{Gin}{width=0.5\linewidth,keepaspectratio}
  \incfig{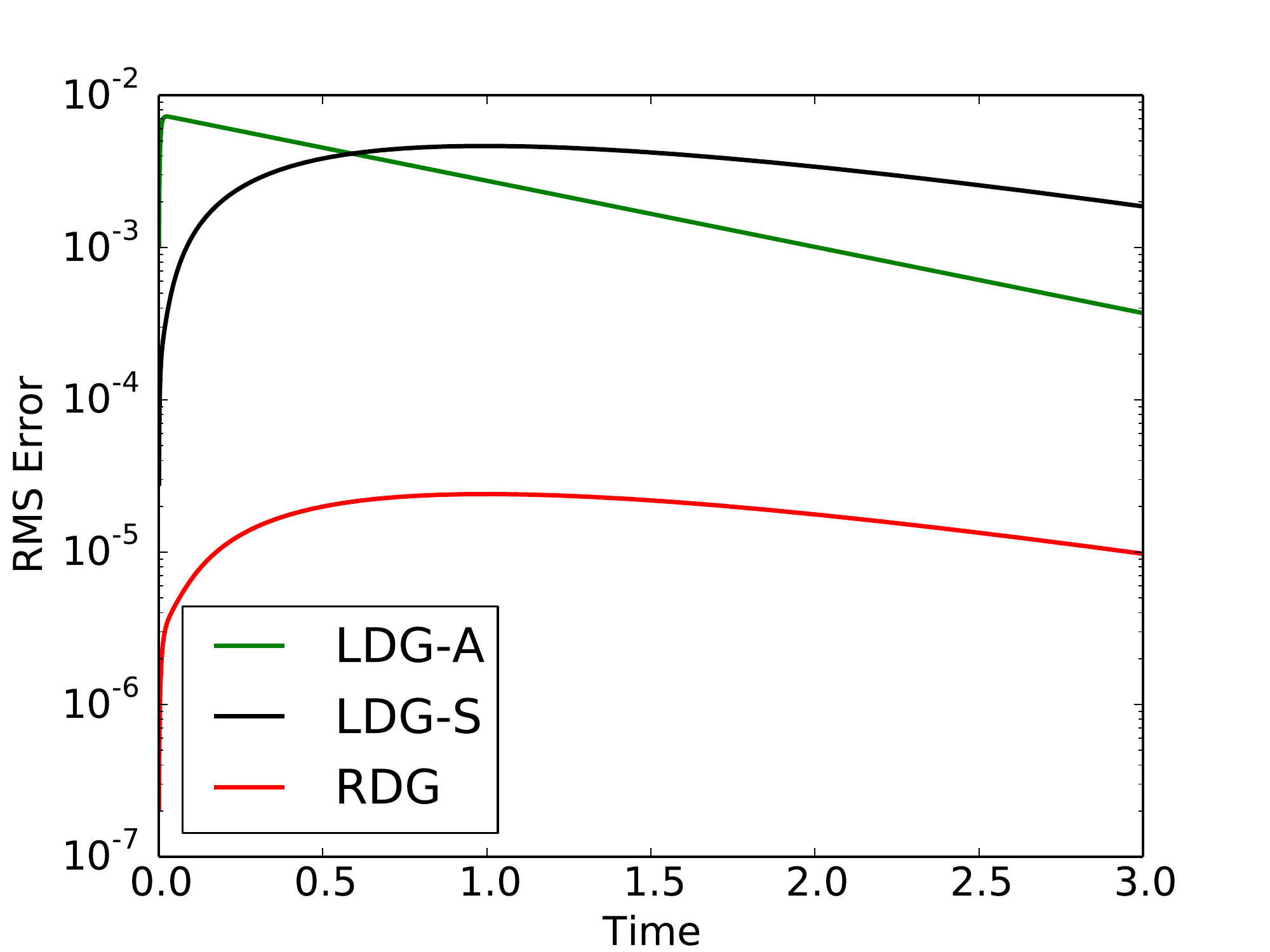}
  \caption{Time history of $L_2$ norm of the error in solution of
    the diffusion equation $f_t=f_{xx}$ on 16 cell mesh with the
    asymmetric LDG scheme (green), the symmetric LDG scheme (black)
    and the RDG scheme (red). The RDG scheme has the smallest
    error. Convergence tests show that the LDG schemes converge as
    $\Delta x^2$, while RDG converges as $\Delta
    x^4$.}\label{fig:err-time-hist}
\end{figure}

Tests of the asymmetric LDG scheme and the RDG are shown in
Figs.\thinspace(\ref{fig:ldg-grid-conv}) and
(\ref{fig:rdg-grid-conv}). These figures show that the LDG scheme,
even though convergent for the solution of a diffusion equation,
mispredicts the slope of $f_{xx}$.

This surprising behavior is confirmed by a Taylor series analysis. To
do this, consider three cells, labeled $j \in \{-1,0,1\}$, centered
around $x=0$. For some function $f(x)$, consider the Taylor expansion
\begin{align}
  f_N(x) = \sum_{n=0}^N \frac{x^n}{n!} f^{(n)}(0),
\label{eq:Taylor-series}
\end{align}
where $f^{(n)}(0)$ is the $n$-th derivative evaluated at
$x=0$. Project $f_N(x)$ on the DG basis functions (Legendre
polynomials, $P_k(\eta_j(x))$, with $\eta_j(x)=2(x-x_j)/\Delta x$,
where $x_j$ is the cell center coordinate) in each cell $j$. Denote
%
these projections by $F_{j,k}$. Use these projections in the stencil,
for example, replacing $T f_k$ by $F_{+1,k}$, $T^{-1}f_k$ by $F_{-1,k}$
and $I f_k$ by $F_{0,k}$, to compute the expansion for the operator
$f_{xx}$. Finally, projecting the resulting expansion back onto a
Taylor basis in cell $j=0$, yields the final Taylor expansion of
$g_h(x)$.

For the asymmetric LDG stencil, \eqr{\ref{eq:1}}, to leading order, the
slope of $g_h(x)$ is $-6 f_{xx}/\Delta x$. Note that the slope should be
$f_{xxx}$, indicating that not only is the slope incorrect, but blows up
as the mesh is refined. This behavior is confirmed quantitatively in
Fig.\thinspace(\ref{fig:ldg-grid-conv}). [Certain types of Taylor series
analysis can give rise to misleading results because of
supra-convergence phenomena, as discussed in
Refs.~\cite{Kreiss:1986,Zhang-Shu:2003}.  The result from the way we use
Taylor series here is confirmed by a von Neumann-like analysis in the
next section.]
For the symmetric LDG scheme,
\eqr{\ref{eq:s-ldg}}, a Taylor series analysis shows that to leading
order the slope is $3f_{xxx}/5$, i.e. although of the correct
derivative order, with the wrong coefficient (which should be
unity). On the other hand, for the RDG stencil, \eqr{\ref{eq:rdg}}, to
leading order the slope is $f_{xxx}$, as it should be. Also, a
convergence study shows that a piecewise linear RDG scheme converges
faster (fourth order accuracy) than the LDG schemes, which are
second-order.

This incorrect behavior of the higher moments (higher than cell
average) of $g_h$ with LDG schemes also occurs with higher order basis
functions. For example, one possible asymmetric LDG stencil with
piecewise quadratic Legendre polynomial basis functions is
\begin{align}
  \left(
    \begin{array}{cc}
      g_{0,j}\\
      g_{1,j}\\
      g_{2,j}\\
    \end{array}
  \right)
  =
  \frac{1}{\Delta x^2}
  \left(
    \begin{array}{ccc}
      9T^{-1}-18I+9T & 7T^{-1}+2I-9T & 3T^{-1}-12I+9T \\
      -27T^{-1}+6I+21T & -21T^{-1}-54I-21T & -9T^{-1}+24I+21T \\
      45T^{-1}-60I+15T & 35T^{-1}+40I-15T & 15T^{-1}-90I+15T
    \end{array}\right)\left(\begin{array}{c}
      f_{0,j} \\
      f_{1,j} \\
      f_{2,j}
    \end{array}
  \right).
  \label{eq:a-ldg-2}
\end{align}

A Taylor series analysis of this stencil shows that, to leading order,
the slope goes as $8f_{xxx}/5$ and the second moment goes as
$-6f_{xxx}/\Delta x$. This shows that not only is the slope
mispredicted, but the highest moment (which should be $f_{xxxx}$)
blows up as $\Delta x \rightarrow 0$. In general, for higher order
asymmetric LDG schemes, all moments are incorrect, with the highest
moment blowing up as $1/\Delta x$. For the piecewise linear scheme,
because the slope goes as $1/\Delta x$, the $L_2$ norm of the error
will not converge to zero as the grid is refined. However, for higher
order polynomial schemes, even though the errors at the cell boundary
are $O(1)$, since the highest moment $p$ goes only as $1/\Delta x$
(and not as $1/\Delta x^p$), the $L_2$ error in the approximation of
$g_h$ still converges to zero as the grid is refined.

\section{Exact time dependence from a von Neumann-like analysis}

The above results can also be demonstrated by calculating the full time
dependence of the LDG solution for a diffusion equation, which can be
done using a von Neumann-like analysis (with a Fourier expansion for the
variations between cells, and an eigenmode analysis for the variations
within a cell), as given in Ref.~\cite{Zhang-Shu:2003}.  (We will follow
their notation in this section.  They solve a diffusion equation of the
form $u_t = u_{xx}$, so their $u$ is equivalent to our $f$.)  They show
how to work out the exact solution to the discretized LDG equations for
$u_{j\pm 1/4}(t)$ ($u$ evaluated at two equally-spaced points in the
$j$'th cell corresponding to LDG with piecewise linear basis functions).
The initial condition is $u(x,t=0) = \sin(k x)$ (and $k=1$).  Using a
Fourier representation for the variation between cells ($u_{j+3/4} =
u_{j-1/4} \exp(ik \Delta x)$, etc.), the solutions for $k \Delta x \ll 1$
are
\begin{align}
u_{j-1/4}(t) = \sin(x_{j-1/4}) \left[ e^{-t} + \frac{(\Delta x)^2}{24}
   \left( e^{-|\lambda_1| t} - e^{-t} \right) \right] + {\cal O}((\Delta x)^3)
\label{eq:u1time}
\end{align}
(this extends to higher order the result in their unnumbered equation
after Eq. 3.22), and
\begin{align}
u_{j+1/4}(t) = \sin(x_{j+1/4}) \left[ e^{-t} - \frac{(\Delta x)^2}{24}
   \left( e^{-|\lambda_1| t} - e^{-t} \right) \right] + {\cal O}((\Delta x)^3).
\label{eq:u2time}
\end{align}
Here, $\lambda_1$ is the first of the two eigenvalues given in their
Eq.~3.18, and is $\lambda_1 \approx - 36 / (\Delta x)^2$ for $k \Delta x
\ll 1$.  The exact solution to the analytic diffusion equation is $u
= \sin(x) e^{-t}$, so one sees from the above two equations that the
errors in $u$ from LDG are bounded for all time and converges like
$(\Delta x)^2$.  However, if we ask what the rate of change of
the solution is, we get
\begin{align}
\frac{d u_{j\pm1/4}}{dt} = \sin(x_{j\pm1/4}) \left[ - e^{-t} \pm \frac{3}{2}
  e^{-|\lambda_1| t} \right] + {\cal O}((\Delta x)).
\end{align}
Since the exact analytical solution is $\partial u(x,t) / \partial t =
u_{xx} = - \sin(x) e^{-t}$, we see that there is an order unity error in
the time-derivative of the solution at early times, but it quickly dies
away in time.  Note that the time-derivative of the cell average,
$(d/dt) (u_{j-1/4}+u_{j+1/4})/2$, converges to the right answer, while
the time-derivative of the slope $(d/dt) (u_{j+1/4}-u_{j-1/4})/(\Delta x
/ 2) = - 6/(\Delta x)$ is the same as we found in the Taylor series
analysis around Eq.~\ref{eq:Taylor-series} above.  This confirms that
there are order unity errors in the discretization of the second
derivative operator, but it also shows that these errors rapidly decay
away in time when used in a diffusion equation, so that LDG converges to
the right answer for diffusion problems.

[Ref.~\cite{Zhang-Shu:2003} evaluates the initial condition $\sin(x)$ at
uniformly spaced grid points, though they point out that the more
accurate thing to do is an $L_2$ projection of the initial condition
onto the DG basis functions.  We have repeated the von Neumann-like
analysis with the more accurate $L_2$ projection, as is used elsewhere
in this paper, and find that it changes some of the higher order terms,
but not the term involving $\exp(-|\lambda_1|t)$ to the order kept here.]

As an aside, we point out that the two eigenvalues given by Eq. 3.18
of Ref.~\cite{Zhang-Shu:2003} are both physically meaningful (in
contrast to some researchers who have called one
eigenvalue/eigenvector as the ``good'' or physical mode, and the
others as ``bad'' or unphysical modes).  Because a Fourier
representation $\exp(i k x_j)$ is used for the between-cell variation,
$k \Delta x \in [0,\pi]$ covers the full range of possibilities.
However, because $\exp(i k x_j)$ is periodic in $k$ when evaluated at
a set of uniformly spaced cell centers $x_j = j \Delta x$, one can
consider $k$ and $k' = k + 2 \pi m / (\Delta x)$ (for integer $m$) as
being equivalent.  The second eigenvalue given by their Eq. 3.18
(which at long wavelengths $k \Delta x \ll 1$ is $\lambda_2 \approx -
k^2$), corresponds to the usual eigenvalue for the $d^2/dx^2$
operator.  The first eigenvalue is the discretized approximation to
the $d^2/dx^2$ operator for an effective wave number $k_{\rm eff} = k
- 2 \pi / \Delta x$, or using a reality condition to consider only
positive $k$, can be taken to be an effective wave number of $k_{\rm
  eff} = 2 \pi / \Delta x - k$.  This explains why $\lambda_1$ scales
as $1/(\Delta x)^2 \sim k^2_{\rm eff}$ for small $k \Delta x$.  In
general, a DG algorithm with $N$'th order polynomial basis functions
will lead to an amplification matrix $G(k,\Delta x)$ (such as in their
Eq. 3.7) of size $(N+1) \times (N+1)$.  This leads to $N+1$
eigenvalues for a given value of $k$ that specifies the $\exp(i k j
\Delta x)$ variation between cells, where $0 \leq k \Delta x \leq
\pi$.  These $N+1$ eigenvalues correspond to different modes
with effective wavenumbers up to $k \Delta x = (N+1) \pi$,
corresponding to the sub-cell variations that can be represented with
higher order DG methods.  (This is illustrated by plots of DG
eigenvalues vs. $k \Delta x$ over the range $[0, (N+1) \pi]$, as shown
in Figs.~1 and 3 of Ref.~\cite{vanLeer:2005kk}, and could be further
illustrated by plots of the eigenfunctions.)

For the piecewise linear case studied here,
even with an accurate $L_2$ projection of the initial condition
$\sin(x)$ onto the DG basis functions, there is still a fraction
$\propto (\Delta x)^2$ of the initial condition put in the high-$k$
eigenmode of the LDG operator, which damps at the very fast rate
$\lambda_1 \propto 1/(\Delta x)^2$, thus creating the order unity
errors in $\partial u/ \partial t = u_{xx}$ at early times.  The RDG
algorithm also has two eigenmodes per $k$, but the projection of the
initial condition onto the RDG high-$k$ eigenmode gives a much
smaller value, so it converges when calculating a second derivative $g =
f_{xx}$.

While the analysis in this section confirms that the error in the second
derivative operator rapidly damps out in time for diffusion problems,
one might wonder what the effect of this error is on problems where a
second derivative term does not correspond to a dissipative effect, such
as in the Schr\"odinger equation or in dispersive wave equations.
Consider for example the normalized Schr\"odinger equation $i \partial
u / \partial t = - u_{xx}$ with the initial condition $u(x,t=0) =
\sin(x)$.  The solution is as given in
Eqs.~\ref{eq:u1time}-\ref{eq:u2time}, but with $t \rightarrow i t$.
Thus the solution for $-u_{xx}$ is
\begin{align}
i \frac{d u_{j\pm1/4}}{dt} = \sin(x_{j\pm1/4}) \left[ e^{-i t} \mp
  \frac{3}{2} 
  e^{-i |\lambda_1| t} \right] + {\cal O}((\Delta x)),
\end{align}
which has an order unity error at any time compared to the exact
solution $i \partial u / \partial t = \sin(x) \exp(-i t)$.
Interestingly, however, an observable like the energy $E = - \langle u^*
u_{xx} \rangle$ still properly converges (here $^*$ denotes a complex
conjugate), as can be seen by noting that $E \propto u_{j-1/4}^* i d
u_{j-1/4}/dt + u_{j+1/4}^* i d u_{j+1/4}/dt$, so the errors from terms
involving $\exp(-i |\lambda_1| t)$ cancel to lowest order.  One might
consider other ways to project initial conditions onto the discrete
basis functions, such as by minimizing the error in a Sobolev norm
instead of an $L_2$ norm.  We leave further investigation of these
issues for the Schr\"odinger equation or dispersive waves to future work.

\section{Discussion and Conclusions}
One may wonder why these errors in the discretized second derivative
operator have not been noticed in the
literature before. The main answer lies in the fact that the net effect
of these errors is small when the
same stencils are used in the solution of parabolic
diffusion or elliptic Poisson equations (which is the steady-state
solution of a diffusion equation).  See
Figs.\thinspace(\ref{fig:ldg-grid-conv}) and
(\ref{fig:rdg-grid-conv}).  That is, the errors in the slopes are
modified in the time-dependent case, and damp out very quickly in time
(or on operator inversion in an elliptic problem), so the solutions
converge for those problems.
This is because the component of $f_h(x)$ with a
large error in the eigenvalue for the discretized operator $\partial^2 /
\partial x^2$ is at small scales of order the grid scale, so this
component damps out
quickly in time.  The time-history of the RMS error of the
solution is shown in Fig.\thinspace(\ref{fig:err-time-hist}).  A
remnant of the difficulty LDG has with $f_{xx}$ can be seen in the
time history, as the error rapidly grows from $0$ to a finite value at
very early times, but then saturates as the component of the solution
with the large error decays away.  
(Table 1 of Ref.~\cite{Cockburn:1998uy} gives the error in the LDG
solution to a diffusion equations at $t=2$, so the rapidly changing transient
error at early times is not noticeable then.)
These errors may be more noticeable in problems where one is
plotting not just $f$ but gradients of $f$ (such as in Navier-Stokes
simulations where one plots the vorticity $\nabla \times {\bf v}$).
Overall, one sees that among
the schemes tested, the RDG scheme has the smallest error. In
addition, it converges faster as the grid is refined.  For problems
involving a hyperdiffusion operator $f_{xxxx}$, if it is evaluated as
two successive applications of an LDG or RDG implementation of a second
derivative operator, the net result will be a stencil that couples 5
adjacent cells.  For RDG with piecewise linear basis functions, it is
possible to directly implement a hyperdiffusion operator with a stencil
that involves only 3 adjacent cells.  (An LDG algorithm that split the
hyperdiffusion operator into 4 coupled first-order equations might be
able to get a similar result with an appropriate choice of the order of
upwinding in various steps.)
%
%

While the full RDG method is higher order accurate, it should be
mentioned that certain properties of a diffusion equation, such as
reducing extrema and thus preserving positivity, are not guaranteed by
higher order methods unless limiters of some sort are applied.
(Reducing extrema is related to not decreasing the entropy $S(t) = -
\int d x f \log(f)$.)  For some applications, simpler second-order
accurate diffusion methods (which can preserve these properties
without additional limiters) may be sufficient.

\vspace{1em}
\noindent {\bf Acknowledgements}
\nopagebreak
\vspace{1em}
\nopagebreak

We thank J.~C.\ Hosea, M.~C.\ Zarnstorff, and S.~C.\ Prager for
supporting the initiation of this project, and we thank B. Cockburn and
C.-W. Shu for helpful discussions.  This work was funded by the
U.~S. Department of Energy Contract DE-AC02-09CH11466, through
the Max-Planck/Princeton Center for Plasma Physics and the Princeton
Plasma Physics Laboratory.

\vspace{1em}
\noindent {\bf References}

\bibliographystyle{elsarticle-num} 
\bibliography{gke.bib}
\end{document}